\documentclass[12pt]{article}
\usepackage{float}
\usepackage[margin=2cm]{geometry}
\usepackage{comment}
\usepackage{amsmath,amssymb,mathtools,graphicx,subfigure,setspace}
\usepackage{cite}
\usepackage{slashed} 		 
\usepackage{color}
\makeatother

\newcommand{\be}{\begin{equation}}
\newcommand{\bea}{\begin{eqnarray}}
\newcommand{\eea}{\end{eqnarray}}
\newcommand{\ba}{\begin{array}}
\newcommand{\ea}{\end{array}}
\newcommand{\ee}{\end{equation}}
\newcommand{\bes}{\begin{equation*}}
\newcommand{\beas}{\begin{eqnarray*}}
\newcommand{\eeas}{\end{eqnarray*}}
\newcommand{\bas}{\begin{array*}}
\newcommand{\eas}{\end{array*}}
\newcommand{\ees}{\end{equation*}}

\numberwithin{equation}{section}
\usepackage{graphicx} 
\usepackage[utf8]{inputenc}
\usepackage[english]{babel}
\usepackage{braket}
\usepackage{xspace}
\usepackage{bbm}
\usepackage{mathdots}
\usepackage{stackrel}
\usepackage[dvipsnames]{xcolor}
\usepackage{braket,physics} 
\usepackage{dsfont}
\usepackage{appendix}
\usepackage{hyperref}
\hypersetup{
 colorlinks=true,
 linkcolor=blue,
 anchorcolor = blue,
 citecolor = blue,
 filecolor = blue,
 urlcolor = blue
}

\begin{document}
\onehalfspacing
\noindent

\begin{titlepage}
\vspace{10mm}
\begin{flushright}
\end{flushright}

\vspace*{20mm}
\begin{center}

{\Large {\bf On Krylov Complexity as a Probe of the Quantum Mpemba Effect
}\\
}

\vspace*{15mm}
\vspace*{1mm}
{Mohsen Alishahiha and  Mohammad Javad Vasli}

 \vspace*{1cm}

{\it { School of Quantum Physics and Matter\\ Institute for Research in Fundamental Sciences (IPM),\\
	P.O. Box 19395-5531, Tehran, Iran\\}  }

 \vspace*{0.5cm}
{E-mails: {\tt alishah@ipm.ir, vasli@ipm.ir}}%

\vspace*{1cm}
\end{center}

\begin{abstract}

We investigate Krylov state complexity as a probe of the quantum Mpemba effect in quantum spin chains. 
For models without global $U(1)$ symmetry, Krylov complexity exhibits clear Mpemba-like crossings, consistent with conventional diagnostics such as the trace distance, while offering a complementary interpretation in terms of Hilbert-space exploration and dynamical delocalization. 
In $U(1)$-symmetric systems, we confirm that the recently proposed symmetric component of Krylov complexity serves as a robust and reliable indicator of the QME, capturing anomalous relaxation even in cases where the total complexity fails to do so.

\end{abstract}

\end{titlepage}

\section{Introduction}

The Mpemba effect refers to the counterintuitive phenomenon in which a system prepared further from equilibrium relaxes faster than one initialized closer to equilibrium. 
Originally observed in water, where initially hotter samples were found to freeze more quickly than colder ones~\cite{Mpemba1969}, the effect has since been identified across a wide range of classical dynamical systems~\cite{{Keller:2018},{Baity:2019},{Kumar:2020},{Schwarzendahl:2022},{Pemartin:2024},{Burridge:2016},{Klich:2019},{Bechhoefer:2021},{Holtzman:2022}}. 
These studies established that the Mpemba effect is not a peculiar feature of water, but rather a generic manifestation of nonequilibrium relaxation, leading to renewed interest in the dynamical mechanisms responsible for anomalous temporal ordering during the approach to equilibrium.  

In recent years, attention has turned toward the quantum regime, giving rise to the so-called quantum Mpemba effect (QME)~\cite{{Joshi:2024sup},{Teza:2025},{Ares:2025},{Yu:2025vth}}. 
In both open and closed quantum systems, it has been observed that states prepared farther from equilibrium may relax faster than those initialized closer to equilibrium. 
In open quantum systems, the effect often parallels its classical counterpart, emerging when overlaps with the slowest Liouvillian modes are suppressed~\cite{{Carollo:2021},{Nava:2019},{Zhang:2025}} (see also~\cite{{Chatterjee:2023},{Moroder:2024},{Nava:2024},{Strachan:2025}}). 
In closed systems, QME-like behavior has been explored through quantum quenches in a variety of models, including spin chains, free fermionic and bosonic systems, random quantum circuits, and many-body localized phases~\cite{{Rylands:2024},{Turkeshi:2024},{Klobas:2024},{Liu:2024},{Yu:2025},{Foligno:2025},{Liu:2024s},{Aditya:2025wzu},{Hallam:2025axu}}.  
These studies demonstrate that the QME is a widespread feature of nonequilibrium quantum dynamics, sensitive to both the nature of the initial state and the spectral properties of the Hamiltonian.

A major challenge in the study of the QME lies in identifying observables that reliably capture it. 
Relaxation processes in quantum systems depend strongly on the interplay between different dynamical channels and on the presence of symmetries or conserved quantities that constrain the evolution. 
Hence, the choice of probe- that is, the quantitative measure of ``distance from equilibrium''- is crucial for revealing Mpemba inversions~\cite{Ares:2025}. 
Among the most informative diagnostics proposed so far is entanglement asymmetry, defined as the difference between the entanglement entropy of a subsystem and that of its symmetry-projected counterpart with respect to a global $U(1)$ charge. 
This measure provides a direct probe of local symmetry restoration and has revealed clear Mpemba-like behavior in both integrable and non-integrable spin chains~\cite{Ares:2022koq,Murciano:2023qrv,Yamashika:2024hpr,Ares:2025ljj,Klobas:2024mlb,Ares:2025vjw}. 
Related analyses have identified Mpemba-type behavior in non-integrable systems without global symmetries, where the effect correlates with spectral indicators such as the inverse participation ratio~\cite{Bhore:2025nko}. 
Together, these developments highlight both the ubiquity of the quantum Mpemba effect and its strong dependence on the probe used to characterize relaxation. 

Recently, Krylov complexity\footnote{For comprehensive reviews, see~\cite{{Rabinovici:2025otw},{Nandy:2024evd}}.} has emerged as a powerful and physically transparent diagnostic of quantum dynamics and chaos~\cite{{Parker:2018yvk},{Dymarsky:2021bjq},{Bhattacharjee:2022vlt},{Avdoshkin:2022xuw},{Camargo:2022rnt},{Vasli:2023syq},{Imani:2025etp},{Rabinovici:2021qqt},{Rabinovici:2022beu},{Scialchi:2023bmw},{Trigueros:2021rwj},{Espanol:2022cqr},{Erdmenger:2023wjg},{Huh:2023jxt},{Camargo:2024deu},{Nandy:2024wwv},{Bhattacharjee:2024yxj},{Balasubramanian:2024ghv},{Baggioli:2024wbz},{Alishahiha:2024vbf},{FarajiAstaneh:2025rlc},{Bhattacharya:2024szw},{Huh:2024ytz},{Baggioli:2025ohh}}. 
By quantifying how rapidly a quantum state or operator explores its accessible Hilbert space, Krylov complexity encodes both the geometric and dynamical aspects of relaxation that are often inaccessible to conventional observables. 
Within this framework, recent work~\cite{Beetar:2025tlf} on a certain spin-chain system with global $U(1)$ symmetry reported clear Mpemba-like behavior in the symmetric (diagonal) component of Krylov complexity, obtained by neglecting contributions from the asymmetric (off-diagonal) sectors. 
Interestingly, inclusion of the full asymmetric contribution tends to smooth out these features, suggesting that coherence between symmetry sectors plays a subtle yet nontrivial role in shaping relaxation hierarchies. 
This observation points toward a deeper connection between complexity growth, symmetry constraints, and the emergence of anomalous relaxation in quantum many-body systems.

In this work, we build upon these developments by systematically investigating Krylov state complexity as a probe of the quantum Mpemba effect in spin chains with and without global symmetries. 
For models without $U(1)$ symmetry, we find that the total Krylov complexity alone exhibits clear and robust Mpemba-like crossings, offering a distinct interpretation in terms of Hilbert-space exploration and dynamical delocalization. 
In contrast, for systems with a global $U(1)$ symmetry, we show that decomposing the complexity into symmetric  and asymmetric components reveals how charge conservation reshapes relaxation dynamics. 
Our analysis demonstrates that, while the total complexity may not always show a pronounced inversion, the symmetric component provides a consistent and reliable diagnostic of Mpemba-like behavior, confirming the proposal of \cite{Beetar:2025tlf}.

The remainder of this paper is organized as follows. 
In Section~\ref{sec:KC-nonsym}, we review the construction of Krylov complexity and analyze its behavior for a mixed-field Ising
chain without global symmetries, demonstrating its sensitivity to Mpemba-like crossings. 
Section~\ref{sec:KC-sym}, focuses on $U(1)$-symmetric systems, where we show that the symmetric component of the complexity serves as a robust probe of the QME. 
The last section is devoted to conclusions.


\section{Krylov Complexity and the Quantum Mpemba Effect without global symmetry}\label{sec:KC-nonsym}

In this section, we explore how Krylov complexity\footnote{Throughout this paper, we examine the Krylov complexity of states, following \cite{Balasubramanian:2022tpr,Alishahiha:2022anw,Caputa:2024vrn,Alishahiha:2022nhe}.} can serve as a probe of the quantum Mpemba effect (QME) in systems without global symmetries. Following \cite{Bhore:2025nko}, we study the mixed-field Ising chain with the Hamiltonian
\begin{equation}\label{eq:mf_ising}
H_{\rm Ising} = \sum_{j=1}^{L-1} \sigma_j^z \sigma_{j+1}^z + g \sum_{j=1}^{L} \sigma_j^x + h \sum_{j=2}^{L-1} \sigma_j^z + h' \sigma_1^z - h' \sigma_L^z,
\end{equation}
where the boundary fields $\pm h'$ explicitly break reflection symmetry, ensuring that no global symmetries remain beyond energy conservation.   

We also examine the next-nearest-neighbor XXZ Hamiltonian~\cite{Ares:2025ljj,Bhore:2025nko}
\begin{equation}\label{eq:XXz2}
H = J_1 \sum_{j=1}^{L-1} \big( \sigma_j^x \sigma_{j+1}^x + \sigma_j^y \sigma_{j+1}^y + \Delta_1 \sigma_j^z \sigma_{j+1}^z \big)
   + J_2 \sum_{j=1}^{L-2} \big( \sigma_j^x \sigma_{j+2}^x + \sigma_j^y \sigma_{j+2}^y + \Delta_2 \sigma_j^z \sigma_{j+2}^z \big)\,,
\end{equation}
which is known to be chaotic and possesses a global $U(1)$ symmetry generated by the total magnetization along the $z$-axis,
\begin{equation}
Q = \sum_{i=1}^{L} \sigma_i^z.
\end{equation}
In addition to the continuous $U(1)$ symmetry, this model also exhibits two discrete symmetries: parity and a global $Z_2$ symmetry. 
The role of these symmetries in shaping Krylov complexity will be explored in more detail in the next section. 
Here, however, we focus on the fully symmetry-broken case in order to isolate the intrinsic dynamical features of complexity growth. 
To this end, we set $J_1 = J_2$ and consider the following Hamiltonian:
\begin{equation}
H = \sum_{i=1}^{L-1} \big( \sigma_i^x \sigma_{i+1}^x + \Delta_y \sigma_i^y \sigma_{i+1}^y + \Delta_z \sigma_i^z \sigma_{i+1}^z \big)
  + \sum_{i=1}^{L-2} \big( \sigma_i^x \sigma_{i+2}^x + \Delta_y \sigma_i^y \sigma_{i+2}^y + \Delta_z \sigma_i^z \sigma_{i+2}^z \big)
  + h \sum_{i=2}^{L-1} \sigma_i^z + h' (\sigma_1^z - \sigma_L^z).
\end{equation}
Here, $\Delta_y \neq 1$ explicitly breaks the global $U(1)$ symmetry, while the boundary and linear field terms break the remaining discrete symmetries, ensuring that the system is fully non-symmetric.

Our goal is to compute Krylov complexity for various initial states and assess its effectiveness as a diagnostic of the QME. The Krylov basis for a normalized state $|\psi\rangle$ is constructed via the Lanczos algorithm \cite{Lanczos:1950zz,viswanath2008recursion}, beginning with $|0\rangle = |\psi\rangle$ and generating subsequent orthogonal vectors recursively as
\begin{equation}\label{GS}
(H - a_n)|n\rangle = b_{n+1} |n+1\rangle + b_n |n-1\rangle,
\end{equation}
with coefficients $a_n = \langle n| H | n \rangle$, $b_{n+1}^2 = \langle n|H^2|n\rangle - a_n^2 - b_n^2$, and $b_0=0$. This procedure produces an orthonormal basis $\{|n\rangle\}$ of dimension ${\cal D}_\psi \leq {\cal D}$, along with the Lanczos coefficients $\{a_n, b_n\}$. Expanding the time-evolved state in this basis,
\begin{equation}\label{phi}
|\psi(t)\rangle = \sum_{n=0}^{{\cal D}_\psi-1} \phi_n(t) |n\rangle, \qquad \sum_{n=0}^{{\cal D}_\psi-1} |\phi_n(t)|^2 = 1,
\end{equation}
the amplitudes $\phi_n(t)$ satisfy the tridiagonal Schr\"odinger equation
\begin{equation}\label{Sch-phi}
i \partial_t \phi_n(t) = a_n \phi_n(t) + b_n \phi_{n-1}(t) + b_{n+1} \phi_{n+1}(t),
\end{equation}
with $\phi_n(0) = \delta_{n0}$. The Krylov complexity is then defined as the expectation value of the number operator \cite{Parker:2018yvk},
\begin{equation}\label{KC-def}
{\cal C}(t) = \langle \psi(t) | {\cal N} | \psi(t) \rangle, \qquad {\cal N} = \sum_{n=0}^{{\cal D}_\psi-1} n |n\rangle \langle n|.
\end{equation}

We consider spin-coherent product states as initial conditions, parameterized by Bloch angles $(\theta,\phi)$,
\begin{equation}\label{initial}
|\theta, \phi\rangle = \bigotimes_{i=1}^L \left( \cos \frac{\theta}{2} |Z+\rangle_i + e^{i\phi} \sin \frac{\theta}{2} |Z-\rangle_i \right),
\end{equation}
where $|Z\pm\rangle_i$ are eigenstates of $\sigma^z_i$. Setting $\phi=0$, the angle $\theta$ serves as a tilt parameter controlling the distance from equilibrium. This reproduces the family of initial states used in \cite{Bhore:2025nko}, where the QME was diagnosed using trace distance and inverse participation ratio (IPR).  

\begin{figure}[h]
\begin{center}
     \includegraphics[width=0.46\textwidth]{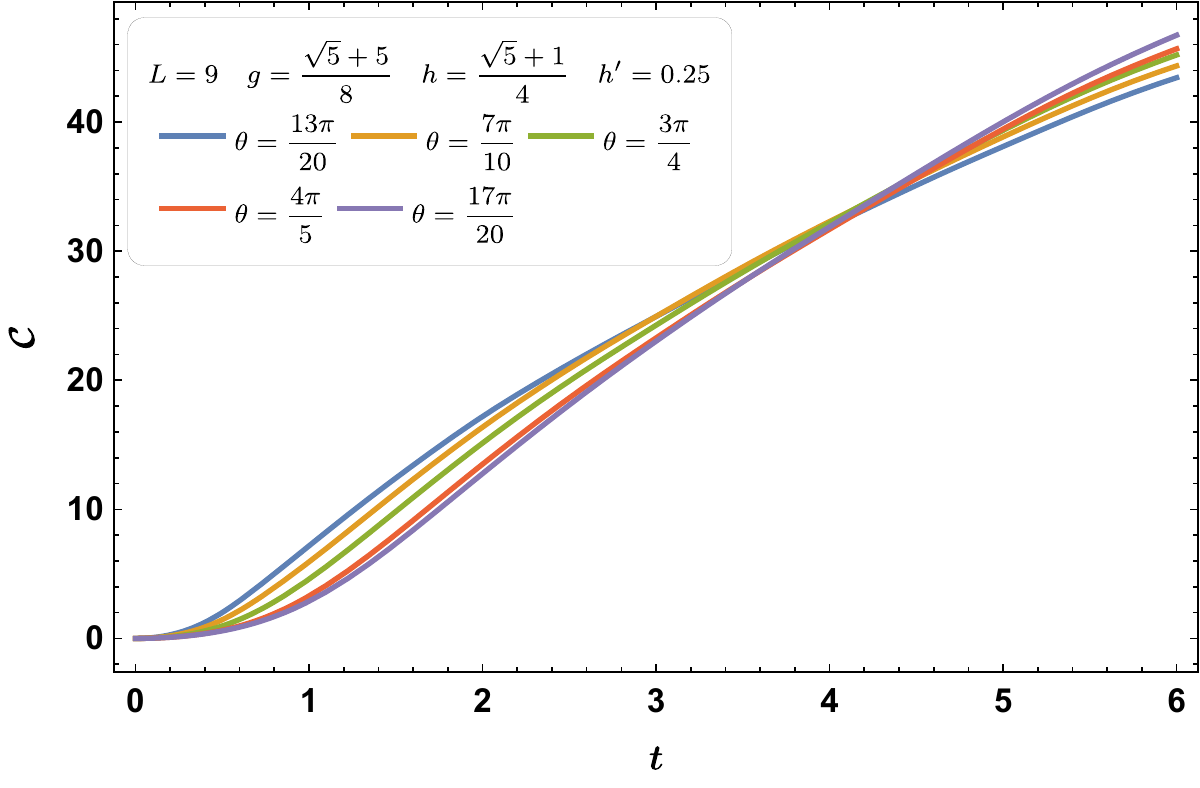}
      \includegraphics[width=0.46\textwidth]{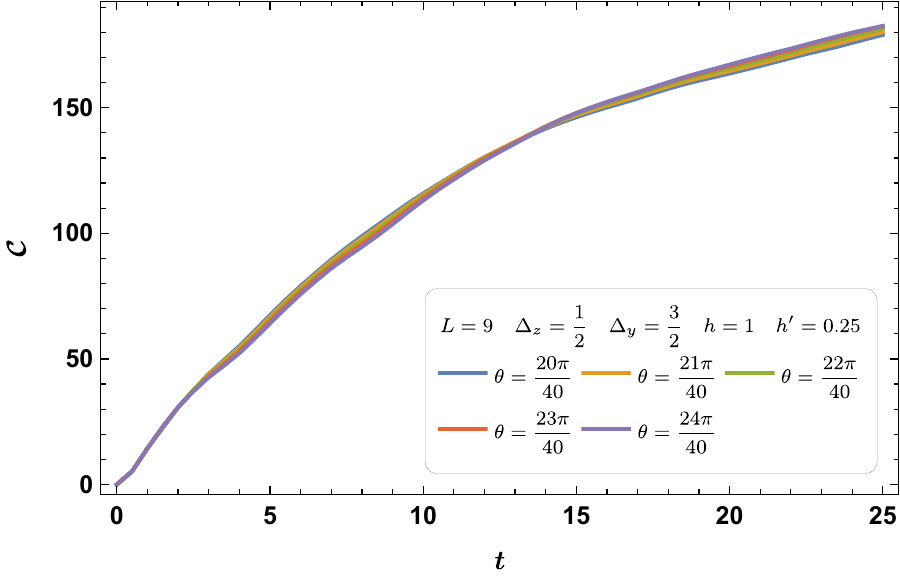}
\caption{Early-time growth of Krylov complexity for different tilt angles $\theta$ in next-nearest-neighbor XXZ Hamiltonian (Right) and mixed-field Ising chain (Left). Strongly tilted states initially grow more slowly but eventually overtake weakly tilted ones, a signature of Mpemba-like behavior.} 
\label{C1}
\end{center}
\end{figure}

The numerical results are shown in Figure~\ref{C1}, with the simulation parameters specified in the inset. For larger tilt angles $\theta$, the complexity initially grows more slowly, remaining below that of states prepared with smaller tilts. However, its growth rate is faster, leading to early-time crossings of the complexity curves. At long times, the more strongly tilted states not only overtake the others but also saturate at higher values of complexity. These crossings are clear signatures of Mpemba-like behavior.

To sharpen the discussion, it is useful to present an analytic expression for the Krylov complexity in the energy eigenbasis. Expanding the initial state as
\begin{equation}
|\psi\rangle = \sum_{\alpha=1}^{\cal D} c_\alpha |E_\alpha\rangle,
\end{equation}
with $|E_\alpha\rangle$ the eigenstates of the Hamiltonian, one finds that the complexity naturally splits into diagonal and off-diagonal contributions\footnote{This decomposition can be interpreted as separating the complexity into a symmetric (diagonal) and an asymmetric (off-diagonal) part with respect to the projection onto the energy eigenbasis.
},
\begin{equation}
    {\cal C}(t)=\sum_{\alpha=1}^{\cal D}
    |c_\alpha|^2 {\cal N}_{\alpha\alpha}
    +\sum_{\alpha\neq \beta}^{\cal D}
    c_\alpha^* c_\beta \, {\cal N}_{\alpha\beta}\, e^{-i\omega_{\alpha\beta}t}, 
    \qquad \omega_{\alpha\beta}=E_\alpha-E_\beta,
\end{equation}
where ${\cal N}_{\alpha\beta}=\langle E_\alpha|{\cal N}|E_\beta\rangle$. The infinite-time average reduces to the diagonal ensemble,
\begin{equation}
\overline{{\cal C}} = \sum_{\alpha=1}^{\cal D} |c_\alpha|^2 {\cal N}_{\alpha\alpha}
= {\rm Tr}(\rho_{\rm DE}{\cal N}), 
\qquad 
\rho_{\rm DE}=\sum_{\alpha=1}^{{\cal D}} |c_\alpha|^2 |E_\alpha\rangle\langle E_\alpha|\,,
\end{equation}
and since ${\cal C}(0)=0$ one finds the constraint
\begin{equation}\label{CoE}
   {\rm Tr}(\rho_{\rm DE}{\cal N})
   =-  \sum_{\alpha\neq \beta}^{\cal D}
    c_\alpha^* c_\beta \, {\cal N}_{\alpha\beta} > 0\,.
\end{equation}
Thus, the saturation value of complexity is controlled by the diagonal part, while the transient dynamics originates from the off-diagonal terms, which eventually cancel at late times due to dephasing. Equation~\eqref{CoE} further shows that a larger saturation value requires stronger initial coherence among energy eigenstates, which also drives a faster initial growth of complexity.

The above analytic analysis shows that such inversions are not guaranteed; instead, they depend sensitively on the distribution of energy eigenstates and on their overlaps with the initial state. Indeed, at early times, precisely the regime where curve crossings are most likely to occur, the Krylov complexity exhibits a universal quadratic growth,  
\begin{equation}
    {\cal C}(t\!\sim\!0)\;\approx\; -\frac{t^2}{8}\,
    \sum_{\alpha \neq \beta}
    \omega_{\alpha\beta}^2 \, c_\alpha^* c_\beta \, 
    {\cal N}_{\alpha\beta}\,,
\end{equation}
where the spectral structure enters explicitly via the squared gaps $\omega_{\alpha\beta}^2$. From this perspective, states with larger long-time saturation values of complexity are those that distribute more broadly across the spectrum
( see also below), thereby sampling a richer set of energy gaps. However, the same broad distribution can suppress the initial growth rate through destructive interference among off-diagonal terms. This suppression at early times, followed by a faster subsequent growth, is precisely the mechanism that allows for Mpemba-like inversions.

\begin{figure}[h]
\begin{center}
    \includegraphics[width=0.46\textwidth]{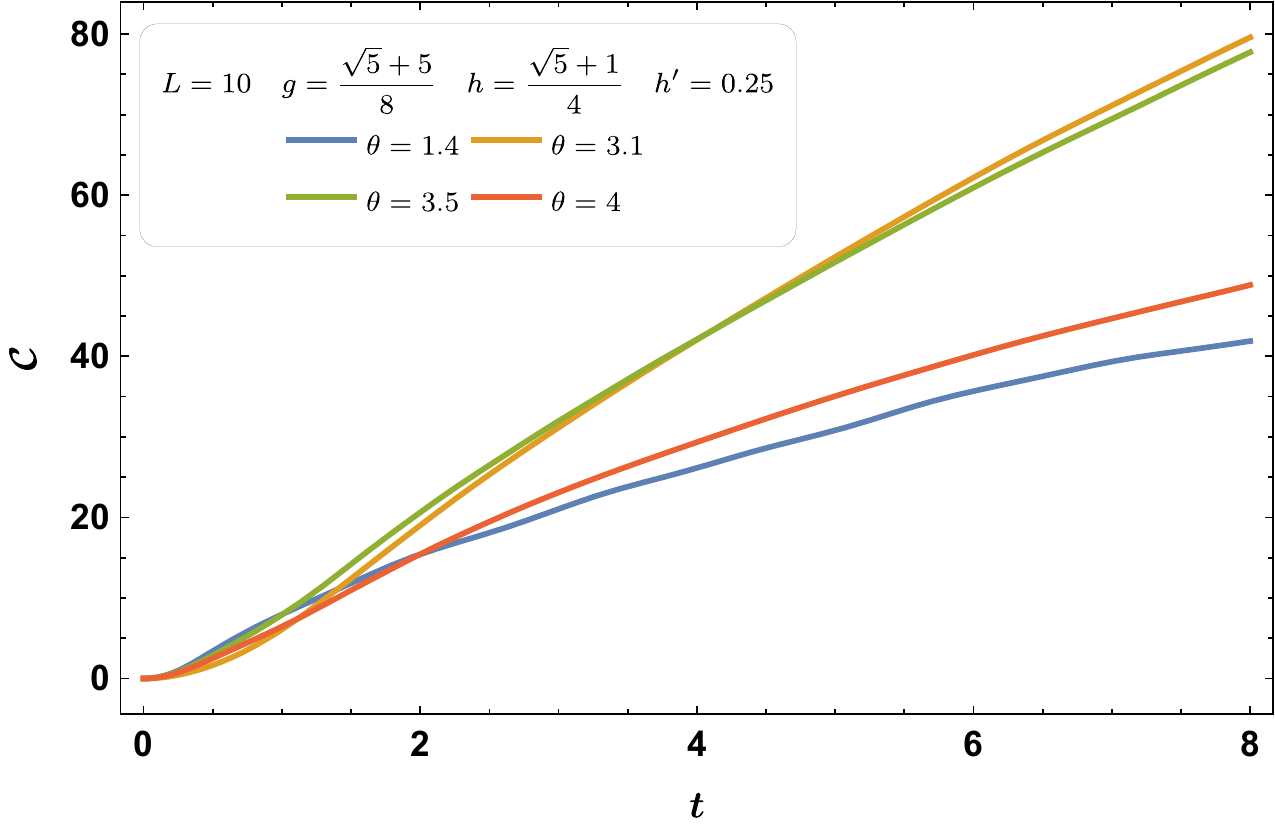}
    \includegraphics[width=0.46\textwidth]{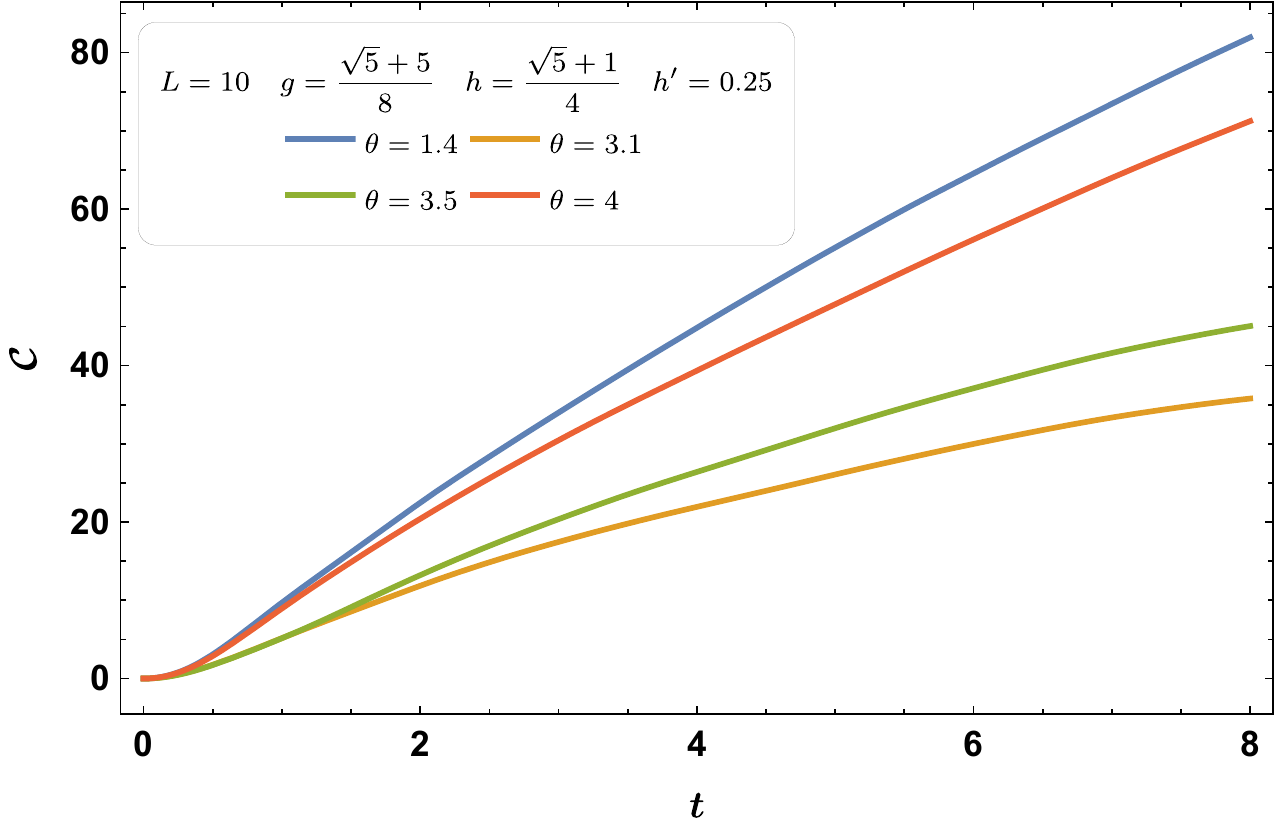}
\caption{Krylov complexity of TFS (Left) and TNS (Right) for different values of $\theta$.} 
\label{C2}
\end{center}
\end{figure}

\begin{figure}[h]
\begin{center}
    \includegraphics[width=0.46\textwidth]{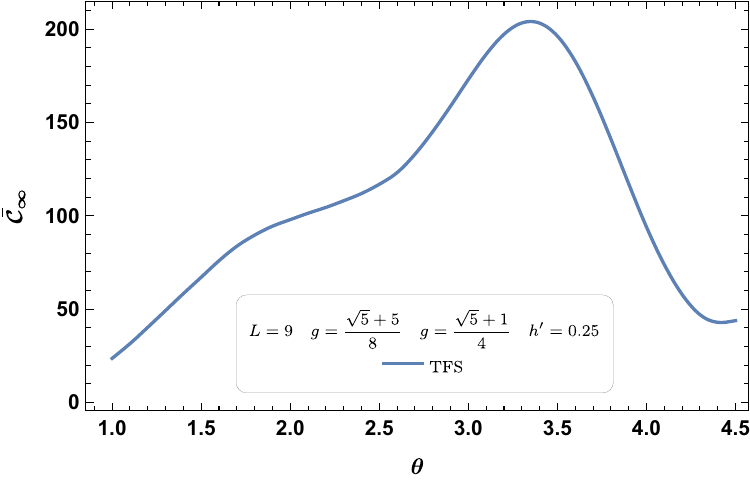}
    \includegraphics[width=0.46\textwidth]{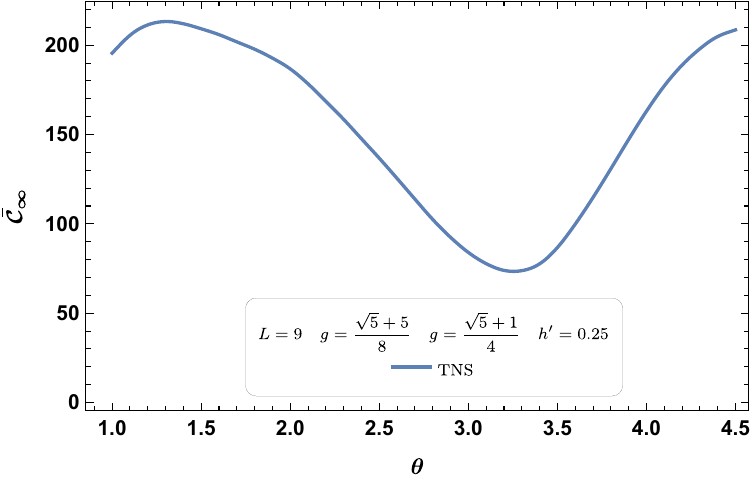}
\caption{Infinite-time averages of Krylov complexity of TFS (Left) and TNS (Right) for different values of $\theta$ and $N=9$.} 
\label{C3}
\end{center}
\end{figure}

To illustrate this point concretely, we compare two classes of initial states: tilted ferromagnetic states (TFS) and tilted N\'eel states (TNS),
\begin{equation}\label{eq:ferro}
    \ket{{\rm F}, \theta}=e^{-i\theta/2\sum_{j=1}^L\sigma_j^y}\ket{\uparrow \cdots \uparrow}, 
    \qquad 
    \ket{{\rm N}, \theta}=e^{-i\theta/2\sum_{j=1}^L\sigma_j^y}\ket{\uparrow \downarrow \cdots \uparrow \downarrow}.
\end{equation}
The corresponding Krylov complexities are shown in Figure~\ref{C2}, and their infinite-time averages in Figure~\ref{C3}.  

The comparison reveals two key insights. First, states with larger $\overline{\cal C}$ grow faster, confirming the correlation between long-time spreading and growth rates. Second, Mpemba-like inversions appear only for TFS: strongly tilted ferromagnetic states initially lag but eventually surpass weakly tilted ones. By contrast, no such inversions are observed for TNS. 
This indicates that while higher saturation values generally correlate with faster growth, this alone is not sufficient to guarantee the occurrence of Mpemba crossings. Indeed, the  Mpemba crossings depend not just on saturation values but also on the microscopic structure of the initial state and its overlaps with the energy spectrum.  

\begin{figure}[h!]
	\begin{center}
		\includegraphics[width=0.4\linewidth]{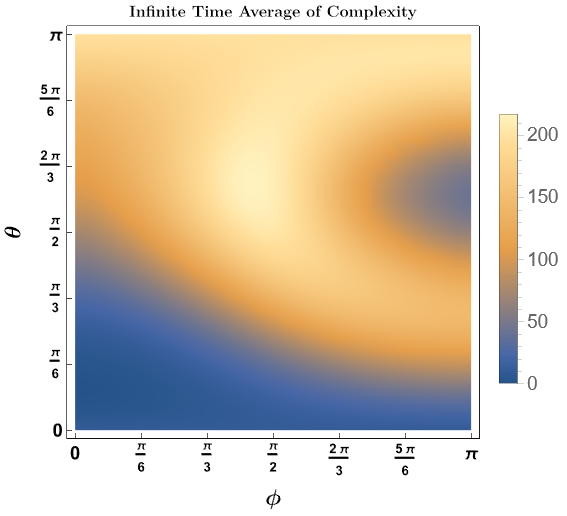}  
		\includegraphics[width=0.4\linewidth]{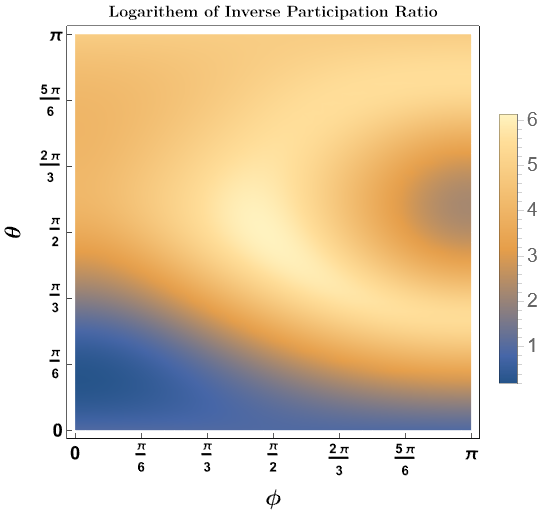}
	\end{center}
	\caption{Infinite-time average of complexity (left) and logarithm of the inverse participation ratio (right) for general initial states given in \eqref{initial}, with $N=9$.  Due to the symmetry under $\phi \rightarrow 2\pi - \phi$, we show the results only for $0 \leq \phi \leq \pi$. }
	\label{fig:complexity}
\end{figure}

Finally, comparing with \cite{Bhore:2025nko}, where QME was identified using trace distance and IPR, we see qualitative agreement: states further from equilibrium relax more rapidly, either via faster decay of trace distance or via earlier crossings in complexity. Moreover, the infinite-time averages of Krylov complexity display near-perfect agreement with the IPR of the initial states, as shown in Figure~\ref{fig:complexity}, reinforcing the view that both quantities capture Hilbert-space delocalization\footnote{This correspondence aligns with previous studies comparing Hilbert-space spreading and IPR in the context of thermalization \cite{Alishahiha:2024rwm}, highlighting that both quantities capture how efficiently a state explores the Hilbert space over time.}.  

However, while these long-time averages align closely, they are not sufficient to diagnose Mpemba-like behavior on their own. The essential signature lies in the early-time dynamics, governed by the interplay of diagonal and off-diagonal contributions. Only by considering both transient growth and asymptotic saturation can Krylov complexity fully capture the nontrivial temporal inversions characteristic of the quantum Mpemba effect.


\section{Krylov Complexity and the Quantum Mpemba Effect in 
$U(1)$-Symmetric Systems}\label{sec:KC-sym}

In this section, we investigate Krylov complexity in quantum systems with a global $U(1)$ symmetry, with the goal of understanding how a conserved charge influences the manifestation of the quantum Mpemba effect.  

When a system possesses a conserved charge $Q$ generated by a global $U(1)$ symmetry, its Hilbert space naturally decomposes into orthogonal subsectors labeled by the charge eigenvalues,  
\begin{equation}\label{Hq}
    {\cal H}=\bigoplus_{q}{\cal H}_q\,,
\end{equation}
where $q$ runs over all possible charge values, and $P_q$ denotes the corresponding projector onto the subsector ${\cal H}_q$.  


Krylov complexity can be formulated in several ways in the presence of symmetry. 
Following the construction introduced earlier, the most direct approach is to evaluate the standard Krylov complexity of a state without explicitly resolving the symmetry structure. 
Given an initial state $|\psi\rangle$, one constructs the Krylov basis $\{|n\rangle\}$ and the associated number operator ${\cal N}=\sum_{n=0}^{{\cal D}_\psi-1}n|n\rangle\langle n|$, yielding the complexity as defined in Eq.~\eqref{KC-def}. 
If the initial state belongs to a particular charge subsector, the entire Krylov basis remains confined within that subsector during the evolution.  

It is often illuminating to express the complexity in a natural energy-charge eigenbasis of the Hilbert space, consisting of the joint eigenstates of $H$ and $Q$,
\begin{equation}
    H|E^{(q)}_\alpha,q\rangle =E^{(q)}_\alpha|E^{(q)}_\alpha,q\rangle,\qquad
    Q|E^{(q)}_\alpha,q\rangle =q|E^{(q)}_\alpha,q\rangle.
\end{equation}
A general normalized state can then be expanded as
\begin{equation}
|\psi\rangle = \sum_{\alpha,q}c_{\alpha q} |E^{(q)}_\alpha,q\rangle ,\qquad \sum_{\alpha,q} |c_{\alpha q}|^2=1,
\end{equation}
where $\alpha$ runs over the dimension ${\cal D}_q$ of the charge sector ${\cal H}_q$. 
Since only a single $U(1)$ symmetry is imposed (and, in particular, no additional discrete symmetry is assumed), the spectra of distinct charge sectors are generally non-degenerate, so that almost all eigenvalues correspond uniquely to a given $q$.  

In this notation, the Krylov complexity of the evolving state can be written as
\begin{equation}\label{Cdecom}
{\cal C}(t) =
\sum_{q,\alpha} |c_{\alpha q}|^2 \, {\cal N}_{\alpha\alpha}^{qq}
+\sum_{q,\alpha\neq \beta} c^*_{\alpha,q}c_{\beta,q} \, {\cal N}_{\alpha\beta}^{qq}\,
     e^{-i\omega_{\alpha\beta}^{qq}t}
+ \sum_{ p\neq q,\alpha,\beta}  c^*_{\alpha,q}c_{\beta,p} \, {\cal N}_{\alpha\beta}^{qp}
    \, e^{-i\omega_{\alpha\beta}^{qp}t},
\end{equation}
which naturally separates into a ``symmetric'' ($q=p$) part (intra-sector contributions) and an ``asymmetric'' ($q\neq p$) part (inter-sector contributions). Here ${\cal N}^{qp}_{\alpha\beta}= \langle q,E^{(q)}_\alpha|{\cal N}|E^{(p)}_\beta,p\rangle$ and $\omega_{\alpha\beta}^{qp}=E^{(q)}_\alpha-E^{(p)}_\beta$.  

Alternatively, one can construct a symmetry-adapted version of Krylov complexity by projecting onto individual charge sectors. 
Defining the subsector number operator
\begin{equation}
{\cal N}_q=\sum_{n=0}^{{\cal D}_\psi-1}n\,P_q|n\rangle\langle n|P_q,
\end{equation}
the symmetric Krylov complexity is given by~\cite{Beetar:2025tlf}\footnote{See also the related notion of symmetry-resolved Krylov complexity introduced in Refs.~\cite{Caputa:2025mii,Caputa:2025ozd}, where the total complexity is expressed as an average over contributions from individual symmetry subsectors.}
\begin{equation}
{\cal C}_S(t)=\sum_q\langle \psi(t)|{\cal N}_q|\psi(t)\rangle
=\sum_{q,\alpha} |c_{\alpha q}|^2 \, {\cal N}_{\alpha\alpha}^{qq}
+\sum_{q,\alpha\neq \beta} c^*_{\alpha,q}c_{\beta,q} \, {\cal N}_{\alpha\beta}^{qq}\,
     e^{-i\omega_{\alpha\beta}^{qq}t}\,,
\end{equation}
which corresponds to the diagonal part of the full decomposition~\eqref{Cdecom}. 
The asymmetric contribution is then
\begin{equation}
   {\cal C}_A(t)= \sum_{ p\neq q,\alpha,\beta}  c^*_{\alpha,q}c_{\beta,p} \, {\cal N}_{\alpha\beta}^{qp}
    \, e^{-i\omega_{\alpha\beta}^{qp}t}\,.
\end{equation}

As in the symmetryless case, the infinite-time average reduces to the diagonal ensemble,
\begin{equation}
    \overline{{\cal C}}=\sum_{q,\alpha} |c_{\alpha,q}|^2 \, {\cal N}_{\alpha\alpha}^{qq}
    ={\rm Tr}(\rho_{\rm DE}{\cal N}),\qquad
    \rho_{\rm DE}=\sum_{q,\alpha}|c_{\alpha,q}|^2\, |E^{(q)}_\alpha,q\rangle\langle q,E^{(q)}_\alpha|\,,
\end{equation}
while the condition ${\cal C}(0)=0$ implies
\begin{equation}\label{CoEq}
  \overline{{\cal C}}=-\left(  \sum_{q,\alpha\neq\beta}c^*_{\alpha,q}c_{\beta,q} \, {\cal N}_{\alpha\beta}^{qq}
  +\sum_{q\neq p,\alpha,\beta}c^*_{\alpha,q}c_{\beta,p} \, {\cal N}_{\alpha\beta}^{qp}\right)>0.
\end{equation}

From these expressions, several important insights emerge. 
First, the saturation value of Krylov complexity is primarily determined by the diagonal contributions from the symmetric part, whereas the transient or pre-saturation dynamics are governed by the off-diagonal components, both intra- and inter-sector. 
Equation~\eqref{CoEq} further implies that achieving a larger saturation value requires stronger coherence among energy eigenstates, which simultaneously enhances the initial rate of complexity growth. 
In systems with a global $U(1)$ symmetry, this coherence arises from two distinct sources: correlations among eigenstates within a charge subsector and interference between different charge sectors.

It follows that focusing exclusively on either the symmetric (diagonal) or the asymmetric (off-diagonal) contribution might be insufficient to capture the full dynamics of Krylov complexity. 
While the asymmetric part exhibits oscillatory behavior that decays through dephasing at long times, it can still encode short-lived coherence effects that strongly influence the transient regime. 
These interference effects can either enhance or suppress early-time features, and are precisely those that give rise to Mpemba-like crossings in the complexity growth curves. 
Neglecting either component therefore could remove essential information about the interplay between coherence and dephasing. 
Consequently, if Krylov complexity is to serve as a faithful probe of relaxation and Mpemba-like behavior, it should, in principle, be considered as a total quantity that integrates both contributions.

Nevertheless, the behavior of complexity depends sensitively on the choice of initial state. 
For certain states, deviations from this general expectation can occur: the transient dynamics may be dominated by intra-sector coherence, leading the symmetric component alone to capture most of the relevant relaxation behavior. 
In these cases, the symmetric Krylov complexity not only governs the asymptotic saturation value but also encodes significant information about the entire dynamical evolution.


This possibility was explicitly demonstrated in Ref.~\cite{Beetar:2025tlf}
(see also \cite{Hetenyi1,Hetenyi2}) for the Aubry–André model~\cite{Aubry:1980}, where the total complexity showed only a qualitative late-time separation, while the symmetric component exhibited robust Mpemba-like crossings, making it a more sensitive diagnostic of the QME. 
In particular, a modified definition of the symmetric complexity was introduced,
\begin{equation}\label{masterC}
   \tilde{{\cal C}}_S(t) = {\cal C}_S(t) +{\cal C}_A(0)\,,
\end{equation}
which effectively subtracts the asymmetric contribution at $t=0$. 
Including the full asymmetric part was found to wash out the crossings, suggesting that the symmetric component, properly normalized, provides a sharper signature of the effect.

\begin{figure}[h!]
\begin{center}
\includegraphics[width=0.46\textwidth]{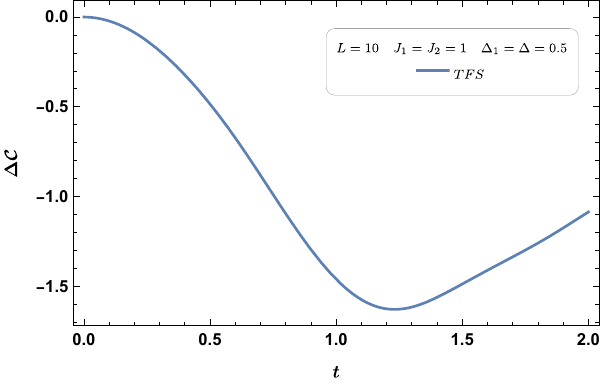}
    \includegraphics[width=0.46\textwidth]{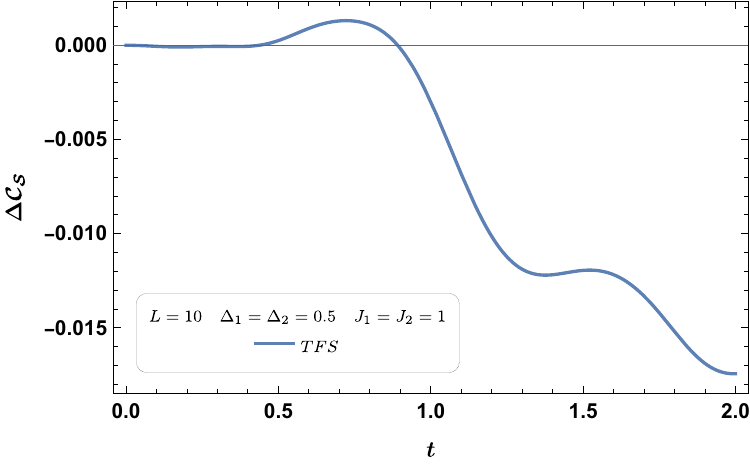}
\caption{Difference in Krylov complexity (left) and its symmetric 
component (right) between initial states with tilt parameters $\theta_1 = 0.15$ and $\theta_2 = 0.2$, for the TFS in the model given by~\eqref{eq:XXz2}. 
System size is $N=10$.} 
\label{H3}
\end{center}
\end{figure}

To examine whether this behavior persists in interacting systems, we now turn to other $U(1)$-symmetric spin chains. 
We first consider the next-nearest-neighbor XXZ 
Hamiltonian \eqref{eq:XXz2}.
To analyze the dynamics, we compute the total and symmetric Krylov complexities for TFS with different tilt parameters. 
To highlight the crossings, we study the difference
\begin{equation}
    \Delta {\cal C}(t) = {\cal C}_{\theta_1}(t) - {\cal C}_{\theta_2}(t)\,,
\end{equation}
and similarly for the symmetric component. 
Using parameters consistent with Ref.~\cite{Bhore:2025nko}, we present results for tilt angles $\theta_1=0.15$ and $\theta_2=0.2$. 
The corresponding results are shown in Fig.~\ref{H3}, where we have also accounted for the model’s additional $Z_2$ symmetry when computing the symmetric complexity.

As seen in Fig.~\ref{H3}, the total complexity does not display a clear Mpemba-like inversion, whereas the symmetric complexity exhibits pronounced crossings, in agreement with the findings of Refs.~\cite{Beetar:2025tlf}. 
This confirms that, in such systems, the symmetric component provides a more reliable probe of the QME, as proposed by \cite{Bhore:2025nko}. 
In contrast, for the TNS, neither the total nor the symmetric complexities show any inversion, consistent with its incoherent relaxation dynamics.

\begin{figure}[h!]
\begin{center}
\includegraphics[width=0.46\textwidth]{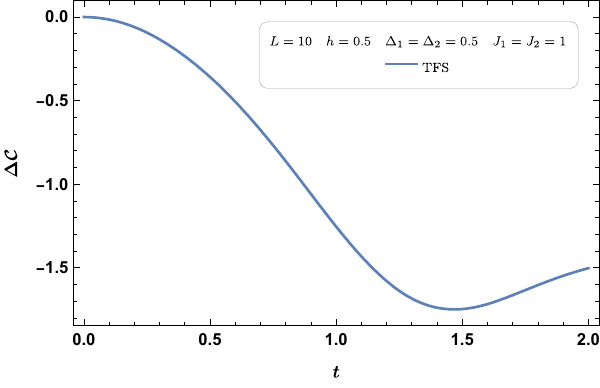}
\includegraphics[width=0.46\textwidth]{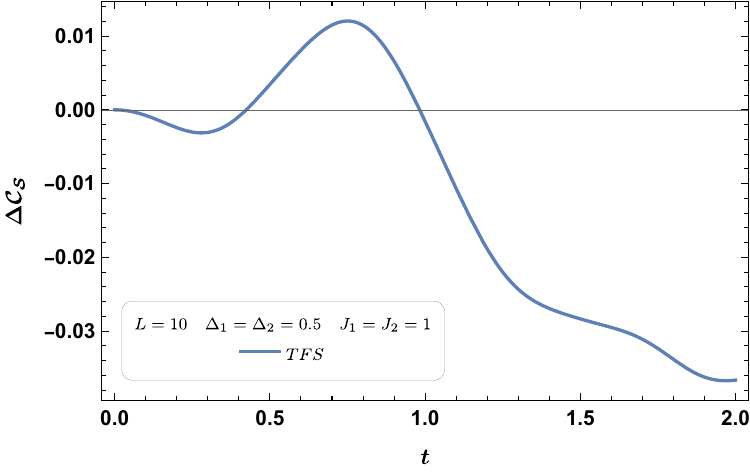}
\caption{Difference in Krylov complexity (left) and in its symmetric 
component (right) between initial states with tilt parameters 
$\theta_1 = 0.15$ and $\theta_2 = 0.2$, for the TFS in the model~\eqref{eq:XXz2} with explicitly broken $Z_2$ symmetry due to the longitudinal field term. 
System size is $N=10$.} 
\label{H3Z2}
\end{center}
\end{figure}

We can also explicitly break the residual $Z_2$ symmetry of the model~\eqref{eq:XXz2} by adding a longitudinal field term of the form 
\begin{equation}
 \frac{h}{2}\sum_{j=1}^L \sigma_j^z,   
\end{equation}
to the Hamiltonian. 
This modification lifts the parity degeneracy and allows us to examine how the breaking of discrete symmetry affects the manifestation of the QME. 
We have repeated the same numerical analysis as before, computing both the total Krylov complexity and its symmetric component for the TFS. 
The corresponding results are presented in Fig.~\ref{H3Z2}.

As shown in Fig.~\ref{H3Z2}, the symmetric component of the complexity exhibits a finer and more structured temporal evolution than the total complexity, reflecting sensitivity to short-time coherence effects that are largely averaged out in the total measure. However, since this behavior is accompanied by significant oscillations and lacks a clear, monotonic inversion, it might not be interpreted as a genuine realization of the QME. Nevertheless, this comparison further underscores that the symmetric component of Krylov complexity can reveal subtle dynamical features of the system.

Let us now consider the integrable XXZ spin chain in a longitudinal field~\cite{Ares:2025ljj},
\begin{equation}\label{eq:xxz}
 H=-\frac{1}{4}\sum_{j=1}^{L-1}\big(\sigma_j^x\sigma_{j+1}^x+\sigma_j^y\sigma_{j+1}^y+\Delta_1 \sigma_j^z\sigma_{j+1}^z\big)
 +\frac{h}{2}\sum_{j=1}^L\sigma_j^z,
\end{equation}
which also possesses a global $U(1)$ symmetry generated by the total magnetization but explicitly breaks the $Z_2$ symmetry through the field term. 

\begin{figure}[h!]
\begin{center}
    \includegraphics[width=0.46\textwidth]{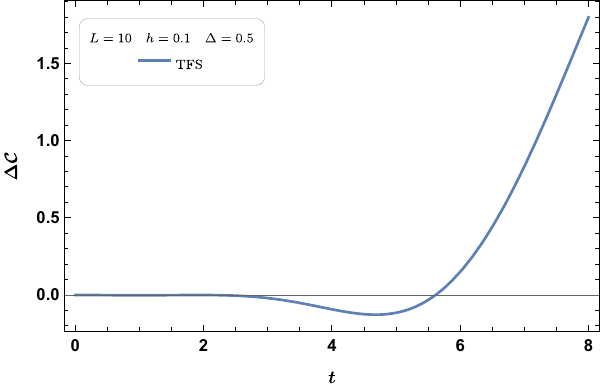}
     \includegraphics[width=0.46\textwidth]{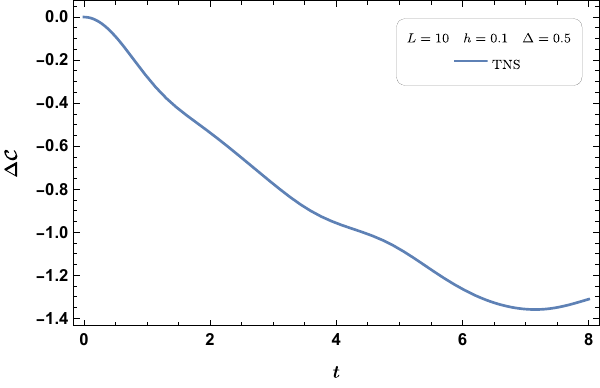}
\caption{Difference in Krylov complexity between initial states with tilt parameters $\theta = 1.2$ and $\theta = 0.8$, for the TFS (left) and for the positive-parity component of the TNS (right). System size is $N=10$.} 
\label{H1}
\end{center}
\end{figure}

For this model, we again analyze the TFS and TNS initial states defined in Eq.~\eqref{eq:ferro}, focusing on tilt parameters $\theta = 1.2$ and $\theta = 0.8$ to compare with Ref.~\cite{Ares:2025ljj}. 
The results are displayed in Fig.~\ref{H1}. 

For the TFS, $\Delta {\cal C}(t)$ changes sign during the evolution, indicating that the corresponding complexity curves cross each other. 
This inversion constitutes a clear signature of the Mpemba-like effect: the state prepared further from equilibrium (larger tilt) initially grows more slowly but eventually overtakes the less tilted state. 
By contrast, for the TNS, $\Delta {\cal C}(t)$ remains of the same sign throughout, showing no evidence of inversion. 
These results agree with those of Ref.~\cite{Ares:2025ljj}, confirming that the full Krylov complexity can indeed capture the short-time reversal of relaxation rates characteristic of the QME in this model.

\begin{figure}[h!]
\begin{center}
    \includegraphics[width=0.46\textwidth]{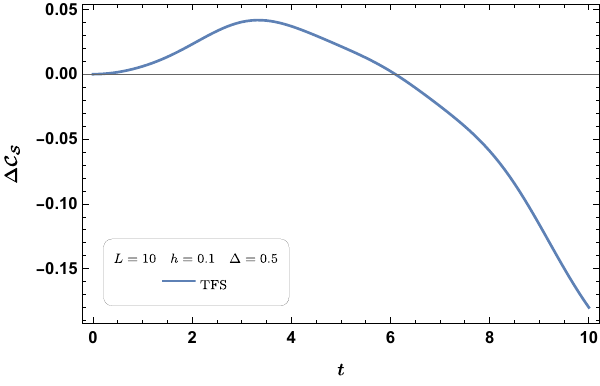}
    \includegraphics[width=0.46\textwidth]{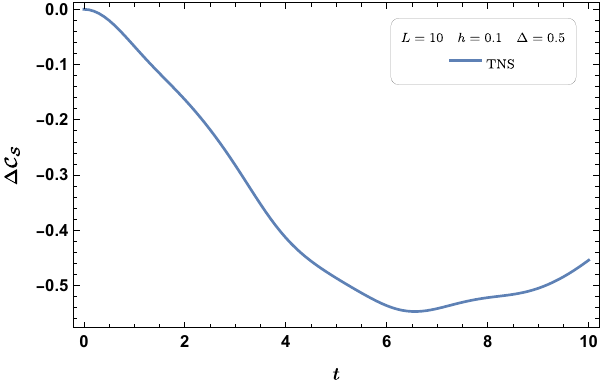}
\caption{Difference in symmetric Krylov complexity between initial states with tilt parameters $\theta = 1.2$ and $\theta = 0.8$, for the TFS (left) and for the positive-parity component of the TNS (right). 
System size is $N=10$.}
\label{SC1}
\end{center}
\end{figure}

To assess the role of symmetry, we further compute the symmetric complexity for both initial states. 
The corresponding results are shown in Fig.~\ref{SC1}. 

From Fig.~\ref{SC1}, we see that for the TFS, the symmetric complexity exhibits a pronounced early-time crossing, while the TNS again shows none. 
This indicates that the essential information about the Mpemba-like inversion is already encoded in the symmetric sector. 
However, unlike the previous case, the infinite-time average of the symmetric complexity does not necessarily increase with tilt. 
The reason lies in the relation
\begin{equation}
   \overline{\tilde{{\cal C}}_S} = \overline{{\cal C}_S} + {\cal C}_A(0)\,,
\end{equation}
which shows that the asymptotic value of the symmetric component depends on the initial coherence contained in the asymmetric sector. 
Thus, while early-time crossings diagnose Mpemba behavior, late-time averages remain insensitive to the initial state.


In summary, our analysis demonstrates that in $U(1)$-symmetric systems, the symmetric component of Krylov complexity provides a robust and reliable probe of the QME. 
Although the total complexity can in some cases reproduce Mpemba-like inversions, these features are generally less pronounced and more sensitive to the choice of initial state. 
The symmetric complexity defined by Eq.~\eqref{masterC}, on the other hand, consistently captures the essential signatures of the effect across all models we have examined, confirming and extending the proposal of Ref.~\cite{Beetar:2025tlf}. 
This finding highlights that Krylov complexity, particularly in its symmetry-adapted form, serves as an insightful diagnostic of anomalous relaxation and the subtle interplay between coherence, dephasing, and symmetry constraints in quantum many-body dynamics.


\section{Conclusions}\label{sec:conclusion}

In this work, we have systematically studied Krylov state complexity as a diagnostic for the QME in both symmetry-broken and $U(1)$-symmetric quantum spin models. 
By analyzing the time evolution of Krylov complexity across different initial states, we have shown that, for systems without symmetry, the structure of complexity growth faithfully encodes the anomalous relaxation patterns that define the QME. 
In particular, the characteristic inversions in relaxation rates—which distinguish the Mpemba effect from conventional thermal equilibration—are directly reflected in the temporal evolution of Krylov complexity. 
This establishes Krylov complexity as a quantitative and physically transparent measure for detecting and characterizing the QME within closed quantum systems.

Turning to models with global $U(1)$ symmetry, our analysis reveals that while the total Krylov complexity can, in certain cases and for specific initial states, display clear Mpemba-like crossings, it does not in general provide a robust probe of anomalous relaxation. 
The total measure is often sensitive to fine details of the initial state and to coherence effects between charge sectors, which can obscure or suppress the inversion behavior that signals the QME. 
A more reliable and universal characterization emerges when the complexity is decomposed into its symmetric and asymmetric components, corresponding respectively to diagonal and off-diagonal contributions in the charge basis. 
By isolating the symmetric component and performing an appropriate shift that compensates for the initial contribution of the asymmetric part, one obtains a quantity that consistently exhibits Mpemba-like behavior across all models studied. 
This shifted symmetric Krylov complexity captures the essential dynamical information governing relaxation, while remaining insensitive to transient oscillations that can mask the effect in the total complexity. 
Our numerical analysis across multiple spin-chain models—both integrable and non-integrable—confirms that this measure provides a stable and physically transparent probe of the QME, thereby supporting and extending the proposal of Ref.~\cite{Beetar:2025tlf}.

These findings demonstrate that symmetry resolution, far from being a mere technical refinement, plays a crucial role in diagnosing anomalous relaxation through complexity. 
In the presence of conserved charges, the symmetric sector encodes the irreversible aspects of Hilbert-space exploration, while the asymmetric part captures transient coherence and dephasing. 
The shifted symmetric complexity effectively combines these contributions into a single measure that reflects the true dynamical hierarchy of relaxation times.

From a broader perspective, our results highlight the potential of Krylov-space methods as powerful tools for investigating nonequilibrium quantum dynamics. 
By linking complexity growth to relaxation pathways and coherence structure, they provide a unifying framework for understanding anomalous thermalization and information scrambling (see, for example, Ref.~\cite{Alishahiha:2024rwm}). 
The sensitivity of Krylov complexity to coherent interference, spectral structure, and the interplay between diagonal and off-diagonal contributions makes it a natural bridge between state  spreading, quantum chaos, and anomalous relaxation phenomena. 
In the context of the QME, it offers a unified language that connects dynamical reversals of relaxation rates to the microscopic geometry of Hilbert-space exploration.

In this work, we have focused exclusively on finite-dimensional many-body systems. 
An interesting extension would be to investigate infinite-dimensional cases, such as quantum field theories, where Krylov complexity may not saturate but continues to grow indefinitely with time. 
In such systems, one can meaningfully study the rate of complexity growth at late times and compare it with the early-time behavior to identify possible inversions. 
Here, the Mpemba-like effect would no longer be manifested through relaxation rates but through the dynamics of complexity growth itself: states that exhibit faster early-time growth may subsequently evolve more slowly at late times. 
In other words, states that initially appear ``more complex'' can eventually become ``less complex'' —a phenomenon that naturally extends the notion of the Mpemba effect to the realm of dynamical complexity.

Taken together, our results in spin-chain models suggest that Krylov complexity provides a unified and physically transparent framework for characterizing Mpemba-like inversions across both finite and infinite-dimensional quantum systems. 
It thus opens new avenues for exploring the interplay between coherence, chaos, and relaxation, and for deepening our understanding of how complex quantum systems approach equilibrium in fundamentally nontrivial ways.


\section*{Acknowledgements}

We would like to thank Mohammad Reza Tanhayi for useful discussions. M.A. would also like to thank the CERN Department of Theoretical Physics for their hospitality during the course of this work. 
The work of M. J. V. is based on research funded by the Iran National Science Foundation (INSF) under Project No. 4047915. We also acknowledge the assistance of ChatGPT for help with text editing and polishing.


\end{document}